\documentclass[12pt]{article}
\usepackage{a4,verbatim,setspace,amsmath}
\usepackage{natbib}
\usepackage{graphicx}
\usepackage{setspace} 

\title{The buckling of a swollen thin gel layer bound to a compliant substrate}

\author{Eric Sultan$^1$ and Arezki Boudaoud$^2$\\
$^1$Instituut-Lorentz, 
P.O. Box 9506, 2300 RA Leiden, The Netherlands.\medskip\\
$^2$Laboratoire de Physique Statistique,\\
UMR8550 du CNRS/ENS/Paris VI/Paris VII,\\
24 Rue Lhomond, 75231 Paris Cedex 05, France.}

\date{\today}

\begin{document}

\maketitle

\centerline{\textbf{Abstract}}
Gels are used to design bilayered structures with high residual stresses. The swelling of a thin layer on a compliant substrate leads to compressive stresses. The post-buckling of this layer is investigated experimentally; the wavelengths and amplitudes of the resulting modes are measured. A simplified model with a self-avoiding rod on a Winkler foundation is in semi-quantitative agreement with experiments and reproduces the observed cusp-like folds.

\newpage 

\section{Introduction}

The buckling of multilayered structures is a strong limitation in the design of sandwich panels~\cite{allen}. This field has been renewed by experiments aiming at the micro-patterning of surfaces through the buckling of thin films bound to compliant substrates (see e.g.~\cite{Genzer06} for a review). For metal films vapor-deposited on an elastomer~\cite{bowden98}, high compressive residual stresses are generated in the film when the system is cooled, due to the mismatch in thermal expansion coefficients between the metal and the elastomer. Recent theoretical efforts addressed nonlinear post-buckling and  herringbone patterns in metal-capped elastomers~\cite{chen04,maha05,huang05}. Residual stresses are also generated in living tissues when growth occurs inhomogeneously in space~\cite{skalak}. Subsequent instabilities can be investigated within the framework of finite elasticity~\cite{alain_martine_long}. In fact, the buckling of multilayered structures was used to explain convolutions in  brain development~\cite{brain,toro}, the organization of seeds on a flower~\cite{steele} or fingerprints formation~\cite{kucken}. 

Our aim is to address experimentally the post-buckling of a thin film on a compliant substrate in the case of strong residual stresses. As thermal expansion induces only small strains, we were led to use polymeric aqueous gels. They are made of a polymeric network immersed in water; they can absorb more water and swell by a length ratio of up to 10~\cite{toyoichi}; their rate of swelling and elastic moduli can be controlled independently by tuning the chemical composition. Two main geometries were investigated experimentally in previous work (see e.g.~\cite{boudaoud03} for a review). The swelling of a gel layer bonded to a rigid substrate results in a cusped oscillating surface with a wavelength proportional to the thickness of the layer~\cite{tanaka87,rapid_tanaka}. The swelling of a 
a gel plate bonded at the edge to a stiff gel results in buckling with a wavelength proportional to the width of the plate~\cite{thierry}, mimicking the wrinkling of the edge of leaves~\cite{sharon,bas_ar}. 

Here were are concerned with a two-layered gel  with a stiff swelling layer (elastic modulus $E_\mathrm{top}$, thickness $h$) ontop a soft non-swelling thick substrate (modulus $E_\mathrm{subs}$). The buckling of such a gel system was investigated theoretically in~\cite{basu06}~: the amount of swelling determines the residual stress $\sigma$ and the buckling stress $\sigma_\mathrm{c}$ and wavelength $\lambda_\mathrm{c}$ are given by the classical formulas~\cite{allen},
\begin{eqnarray}
\frac{\sigma_\mathrm{c}}{E_\mathrm{top}}=\frac{1}{3^{1/3}}\bigg(\frac{E_\mathrm{subs}}{E_\mathrm{top}}\bigg)^{2/3},\label{scathr}\\
\frac{\lambda_\mathrm{c}}{h}=\frac{2\pi}{3^{1/3}}\bigg(\frac{E_\mathrm{top}}{E_\mathrm{subs}}\bigg)^{1/3},
\label{scaling}
\end{eqnarray}
with a Poisson ration $\nu=1/2$ as gels are in general incompressible. Above threshold, the amplitude $A$ of oscillation of the post-buckled state should be a function of the residual stress $\sigma$~\cite{chen04},
\begin{equation}
\frac{A}{h}=\sqrt{\frac{\sigma}{\sigma_\mathrm{c}}-1}.\label{scampl}
\end{equation}

In the present study, we investigate experimentally post-buckled states in the case of strong residual stresses. Incidentally, the cusped oscillating shapes obtained are reminiscent of brain convolutions~\cite{brain}. The article is organized as follows. We describe our experimental set-up and a simplified model with a compressed self-avoiding rod bound to an elastic foundation.  Then we compare and discuss the experimental and numerical results.

\section{Experiments}

\begin{table}[h]
\begin{center}
\begin{tabular}{l c c}
{} & S (substrate) & T (top layer)\\
\noalign{\smallskip}\hline\noalign{\smallskip}
$[$AA+BISAA$]$ & 720 & 1202\\
BISAA:AA ratio & 1:37.5 & 1:19\\
$[$SA$]$ & 0 & 229\\
Swelling ratio (3D)& 1.06 & 1.5 \\
Swelling ratio (2D)& 1.09 & 1.8 \\
$E$ (Pa) & $5.0\cdot 10^3$ & $1.7\cdot 10^4$ \\
Poisson ratio & $1/2$ & $1/2$
\end{tabular}|
\end{center}
\caption{Composition and properties of the gels mainly used in the experiments. Concentrations are given in mmol.L${}^{-1}$. The swelling ratio corresponds to the length dilation factor of a gel piece after swelling, when it is free (3D) or constrained in plane strain (2D). $E$ is the elastic modulus of the gel (after swelling).}
\label{typesgel}
\end{table}

The principle of the experiments is to first prepare the substrate layer then pour the solution for the thin top layer. We made our gels as in~\cite{toyoichi,tanaka87,rapid_tanaka,thierry}. A mixture of acrylamide (AA) and N,N'-methylenebisacrylamide (BISAA) is dissolved with sodium acrylate (SA) in distilled water. The polymerization is initiated by ammonium persulfate (PA) and is catalysed with N,N,N',N'-tetramethylenediamine (TEMED) (0.3\% in volume).  The composition of the gels  is given in Table~\ref{typesgel}. Once the mixture is completed, gelation (polymerization and solidification of the solution) occurs in a few seconds at room temperature. In order to obtain uniform layers --- especially for thin layers --- the solutions were cooled to slow gelation and allow the liquid to spread completely on the substrate.

The characteristics of the gel can be tuned by varying the concentrations of the components. The more concentrated (and, for a same concentration, the more concentrated in BISAA) the solution, the stiffer the gel. Likewise, the swelling ratio (the ratio between a free gel dimensions before and after swelling) can be increased by adding sodium acrylate. For the purpose of the experiment we prepared two distinct types of gel: (S) a soft and non-swelling gel; (T) a stiff and swelling gel. The elastic and swelling properties of these gels were measured and are reported in Tab.~\ref{typesgel}. In order to obtain a compliant substrate, we chose the highest ratio between the elastic moduli allowed by the experiment.

\begin{figure}
\centering
\includegraphics[width=0.5\textwidth,angle=0]{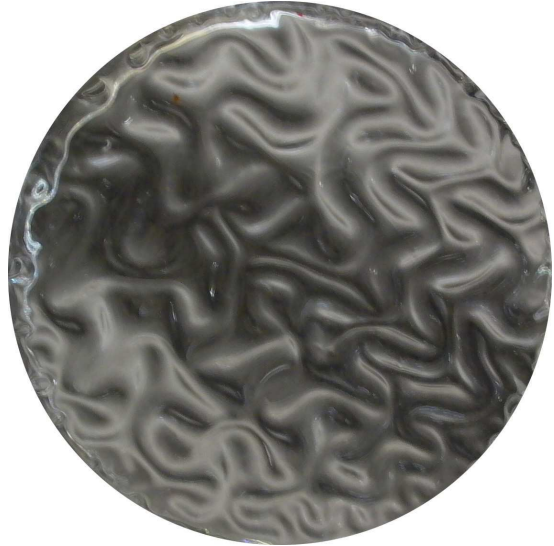}
\caption{Top view after swelling of the top layer (thickness $h=1$ mm above a 3 cm thick substrate) in a dish of 10 cm diameter. The valleys are lighter than the crests. The lighting is not  uniform as a dark strip was placed below to increase the contrast.}
\label{bulk}
\end{figure}

In preliminary experiments, we prepared a thick (3 cm) substrate made of the soft gel (S) in a Petri dish; a 1 mm-thick layer of the second solution (T) was poured above. After gelation, the two layers were chemically bound to each other: no delamination was observed. Then water was added in the dish. It was slowly absorbed by the top layer which started to wrinkle. The wrinkle evolution was slow and a stationary state was reached after half an hour to one hour. A corresponding top view is shown in Fig.~\ref{bulk}. The pattern looks like a superposition of sinusoidal modes with random directions and a wavelength of about 1 cm.
The thickest (1cm) top layers that we used would not reach a stationary state before 2 to 4 days. This 
is consistent with the fact that absorption of water by the gel is a diffusive process~\cite{toyoichi}, so that the equilibration time is proportional to the square of the thickness.

\begin{figure}
\centering
\includegraphics[width=0.7\textwidth,angle=0]{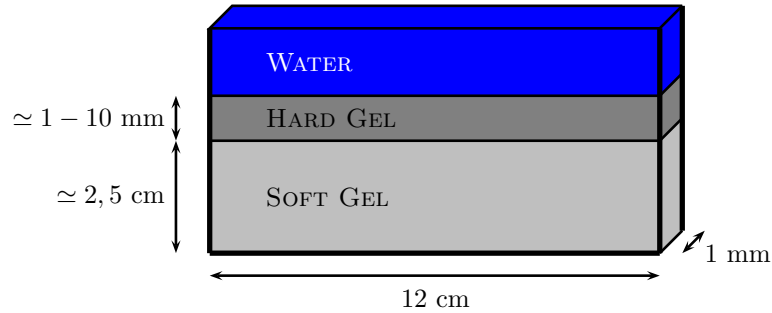}
\caption{Typical geometry and dimensions of the main experimental setup. Plane displacements are constrained by enclosing the whole between two glass plates with a gap of 1 mm.}
\label{geom}
\end{figure}

In order to monitor easily the displacements of the top layer and get a better experimental control, we designed a set-up to constrain plain strain. The cell was made with two glass plates with a 1 mm gap. Rubber stripes were used as spacers and delimited the bottom and sides of the cell. The glass plates were held together by clamps. The gel bi-layer was prepared in the cell according to the geometry depicted in Fig.~\ref{geom}: a hard swelling top (T) with thickness in the 1--6 mm range and a soft substrate (S) with a thickness of about 2.5 cm.

\begin{figure}
\centering
\includegraphics[width=0.8\textwidth,angle=0]{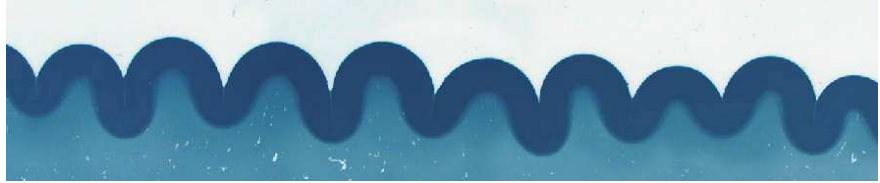}
\caption{Side view of the gel top layer (colored with ink) after swelling. Thickness $h=3$ mm (after swelling), wavelength $\lambda=8.9$ mm and amplitude $A=5.1$ mm. }
\label{equ}
\end{figure}

\begin{figure}
\centering
\includegraphics[width=0.5\textwidth,angle=0]{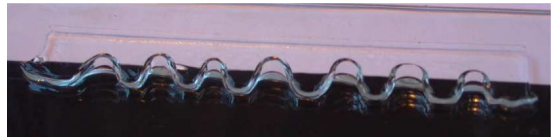}
\caption{View of the gel after taking out one glass plate showing that the cusps did not damage the gel. The swollen layer is oscillating out of plane due to the strong compressive residual stress. }
\label{outof}
\end{figure}

Absorption of water by the upper gel layer initialized swelling.  In the early stages of the experiments, we observe the appearance of fold structures (usually in the center of the cell) separated by a few mm; subsequent appearing structures are then separated by progressively larger and larger distances. Eventually, once water has penetrated across the entire gel layer, the system reaches an equilibrium state for which patterns have well-defined wavelengths and amplitudes of the order of a few mm. The surface of the gel oscillates with cups (Fig.~\ref{equ}). In order to check wether the gel was damaged at these cusps, we opened the cell by taking out one glass plate (Fig.~\ref{outof}); it turns out that no failure occurs at the surface of the top layer.

\section{The model}

\begin{figure}
\centering
\includegraphics[width=0.55\textwidth,angle=0]{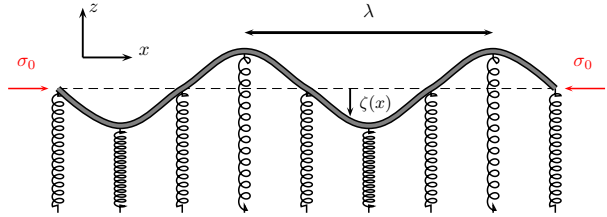}
\caption{Geometry of the model.}
\label{ressorts}
\end{figure}

\begin{figure}
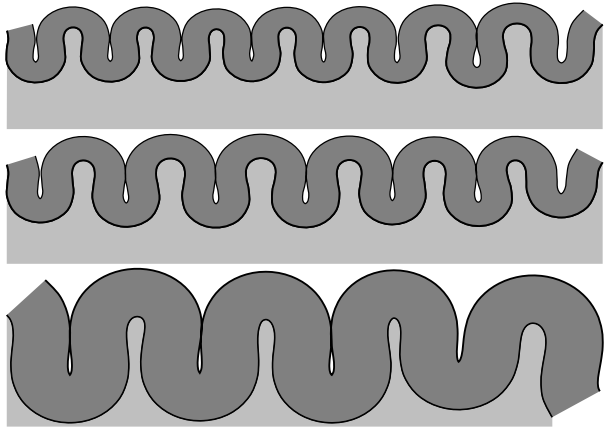

\centering
\includegraphics[width=0.55\textwidth,angle=0]{cusps_00.epsi}\\
\includegraphics[width=0.55\textwidth,angle=0]{cusps_01.epsi}\\
\includegraphics[width=0.55\textwidth,angle=0]{cusps_02.epsi}
\caption{Numerical equilibrium configurations for three values of the thickness $h$ (a) 1.7 mm, (b) 2 mm and (c) 4 mm. The Young modulus of the substrate is $E_\mathrm{subs}=4.6\ 10^4$ Pa and the Young modulus of the top layer $E_\mathrm{top}$ is (a) 292 Pa, (b) 375 Pa and (c) 585 Pas, respectively.}
\label{simusfig}
\end{figure}

In order to interpret the experimental results, we generalize here the classical model of plate/rod on a Winkler foundation, remaining in the framework of linear elasticity. In the following, we will formulate the model for a rod and count the elastic energies per unit of gap of the cell, because the strains are constrained to be plane in the experiment. As we are interested in the post-buckling regime, we assume the rod to be inextensible. Indeed for large displacements of the film, the energetical cost of stretching becomes large compared to the cost of bending~\cite{el_landau}; therefore stretching is avoided and the film may be assumed inextensible. Two important experimental  features should be taken into account. First there is an asymmetry between the two surfaces of the top layer. Second cusps involve self-contact of the upper surface. As a consequence, we study a self-avoiding inextensible rod on a Winkler foundation linked to the lower surface of the rod.
 

More precisely, we consider an elastic rod with neutral line $\mathbf{r}(s)=x(s)\mathbf{e}_x+z(s)\mathbf{e}_z$ parametrised by the arc-length $s\in(0,L)$ (Fig.~\ref{ressorts}). We define the tangential and normal unit vectors by 
$\mathbf{t}(s)=\mathbf{r'}(s)$, $\mathbf{n}(s)=\mathbf{t'}(s)/\|\mathbf{t'}(s)\|$ and the curvature by $\kappa(s)=\mathbf{n}(s)\cdot\mathbf{t'}(s)$. 
In order to take into account the finite thickness $h$ of the gel layer, we construct its lower and upper surfaces by defining $\mathbf{r}^\pm(s)=\mathbf{r}(s)\mp\frac{h}{2}\mathbf{n}(s)$. 

We take as a reference configuration a free rod of length $L$ and thickness $h$ (corresponding to the state after swelling).  Then we match it to a Winkler foundation of length $L_0$, such that $L/L_0>1$ is the swelling ratio. To each configuration $\mathbf{r}(s)$, we associate the elastic energy:
\begin{eqnarray}
\mathcal{E}[\{\mathbf{r}\}]=\frac{1}{2}D\displaystyle\int_0^L\kappa(s)^2\mathrm{d}s+\frac{1}{2}k\int_0^L\left(\mathbf{r}^-(s)-s\tfrac{L_0}{L}\mathbf{e}_x\right)^2\mathrm{d}s\nonumber\\
-\sigma_\mathrm{r} h\left(\mathbf{r}(L)-\mathbf{r}(0)-L_0\mathbf{e}_x \right)\mathbf{\cdot e}_x +\mathcal{H}[\mathbf{r}^+].
\label{nrj}
\end{eqnarray}
The first term is the bending energy with a bending stiffness $D=E_\mathrm{top}h^3/9$, assuming an incompressible material. The second is the energy of the Winkler foundation where $x$ and $y$ displacements of the lower surface of the rod are taken into account -- the reference state $\mathbf{r}^-(s)$ of the foundation has a length $L_0$, so that  $\mathbf{r}^-(s)|_\mathrm{ref}=sL_0/L\mathbf{e}_x$; the stiffness of the foundation is given by $k=4\pi/3\ E_\mathrm{subs}/\lambda$ to be equivalent to an elastic half-space of Young's modulus $E_\mathrm{subs}$ deformed with a wavelength $\lambda$ -- note that in the experiment, the thickness of the substrate is larger than the observed wavelengths so that it is legitimate to consider the substrate as half-infinite. In the numerics, $k$ was fixed and $E_\mathrm{subs}$ deduced from the value of $\lambda$. The third term contains the compressive stress $\sigma_r$, taken as a Lagrange multiplier, needed to achieve the projected length $L_0$ on the $\mathrm{e}_x$ direction. The fourth and last term 
is a purely geometric contribution introduced to forbid self-crossing of the upper interface. More precisely, $\mathcal{H}[\{\mathbf{r}^+\}]=+\infty$ if $\mathbf{r}^+(s)=\mathbf{r}^+(s')$ has at least one solution other then $s= s'$ and $\mathcal{H}[\{\mathbf{r}^+\}]=0$ otherwise.

We performed the study of this rod-like model through the numerical minimization of energy (\ref{nrj}). The rod was descretized in $N$ parts (100--500 in practice), each of length $L/N$, by defining coordinates $\mathbf{r}_i=x_i\mathbf{e}_x+z_i\mathbf{e}_z$ with $x_i=\frac{L}{N}\sum_{j=1}^i\cos\theta_j$ and $z_i=\frac{L}{N}\sum_{j=1}^i\sin\theta_j$ $(i=1,\ldots,N+1)$. Introducing the tangent $\mathbf{t}_i$ and normal $\mathbf{n}_i$ vectors to segment $(\mathbf{r}_i,\mathbf{r}_{i+1})$, the lower and upper interfaces were reconstructed by $\mathbf{r}^\pm_i=\frac{1}{2}(\mathbf{r}_{i}+\mathbf{r}_{i+1}\mp h\mathbf{n}_i)$. The energy is then a function of the $\theta_i$. Powell's algorithm of minimization~\cite{nr}Ê allowed to cope with the discontinuous behavior of $\mathcal{H}[\{\mathbf{r}^+\}]$. Examples of minimal energy configurations are shown in Fig.~\ref{simusfig} with values of the parameters in the experimental range. It can be seen that two experimental features are reproduced: large amplitude oscillations and self-contacting upper surface. A more quantitative comparison with the experiments is the subject of the next section, whereas the limitations of the model are discussed in the conclusion.

\section{Results}

Before discussing the observations on wavelengths and amplitudes, let us estimate the residual stress induced by swelling and compare it to the buckling threshold. The swelling ratio roughly gives the residual strain $\epsilon\simeq0.8$, which induces a residual stress  $\sigma=2/3 E_\mathrm{top}/\epsilon$ assuming plane strain and incompressibility of the material. The ratio of this stress to the threshold is $\sigma/\sigma_c=2 (E_\mathrm{top}/3E_\mathrm{subs})^{2/3}\epsilon\simeq2.2$, so that we expect buckling.

\begin{figure}
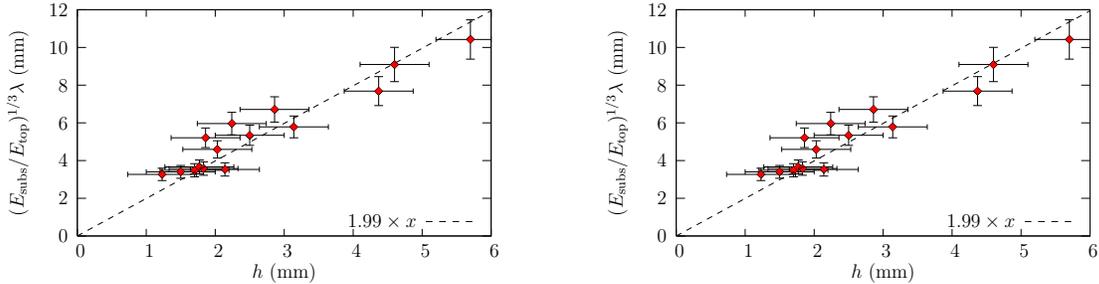

\centering
\includegraphics[width=0.45\textwidth,angle=0]{manips_lo_plot.epsi}\hfill\includegraphics[width=0.45\textwidth,angle=0]{manips_lo_plot.epsi}
\caption{Experiments: Wavelength (left) and amplitude (right) of the gel oscillations as a function of the top layer thickness.  The wavelength $\lambda$ was rescaled by the dimensionless coefficient $(\frac{E_\mathrm{subs}}{E_\mathrm{top}})^{1/3}\simeq0.67$. Horizontal errorbars correspond to the ($\simeq1$ mm) measure error on the thickness $h$; vertical bars correspond to (typically $10\%$) fluctuations on the distance between two adjacent folds and their amplitude.}
\label{manips}
\end{figure}

At constant mechanical properties (same $E_\mathrm{top}$ and $E_\mathrm{subs}$), we measured the equilibrium wavelengths and amplitudes of oscillation. The thicknesses of the upper gel layer was in the range 1--6 mm leading to wavelengths from 6 to 20 mm and amplitudes from 4 to 11 mm (Figure \ref{manips}). The data can be fitted by a linear dependance. We find $\lambda\simeq2({E_\mathrm{top}}/{E_\mathrm{subs}})^{1/3}h$ which is clearly below the classical value $\lambda=4.4({E_\mathrm{top}}/{E_\mathrm{subs}})^{1/3}h$ (Eq. \ref{scaling}). This last value should also hold above threshold in the small slope regime~\cite{chen04}. Similarly, we find $A\simeq\ 1.9h$ which is above the value $A=1.1\ h$ given by Eq.~(\ref{scampl}). 

We also modified Young's modulus of the gels as allowed by the chemical composition. Decreasing their ratio yielded only a small change in wavelengths. This can be accounted for by the weak dependance (power 1/3) on the ratio ${E_\mathrm{top}}/{E_\mathrm{subs}}$. There seemed to be no dependance neither on the thickness of the substrate nor on the gap, which were respectiveley larger and smaller than the wavelength.

\begin{figure}
\centering
\includegraphics[width=0.45\textwidth,angle=0]{simu_lo_plot.epsi}\hfill\includegraphics[width=0.45\textwidth,angle=0]{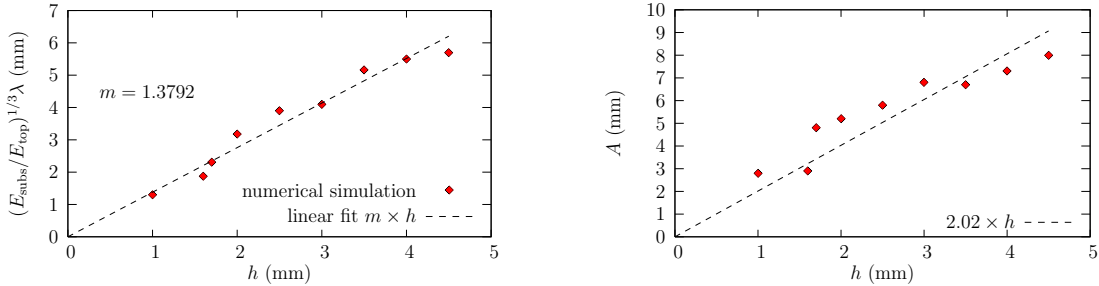}
\caption{Numerics: Wavelength (left) and amplitude of the gel oscillations as a function of the top layer thickness.  The wavelength $\lambda$ was rescaled by the dimensionless coefficient $(\frac{E_\mathrm{top}}{E_\mathrm{subs}})^{1/3}$.}
\label{simus}
\end{figure}

The classical results seem to provide the correct dependance of the wavelength and amplitude but not the prefactors. We turn now to the results of the numerical simulation of the self-avoinding rod on a Winkler foundation as introduced above. Fitting the data to a linear dependance (Fig.~\ref{simus}) yields $\lambda\simeq1.4({E_\mathrm{top}}/{E_\mathrm{subs}})^{1/3}h$ and $A=2.0\  h$, which are much closer to experiments. In fact, accounting for  large slopes tends to shorten wavelengths and increase amplitudes.  Moreover, the simulations also reproduce the qualitative shapes observed in the experiments; the interface between the two gels is smooth whereas the oscillation of the free surface involves self-contacts and cusp-like folds.

\section{Conclusion}
Using the properties of gels, we built a bilayered structure with high residual stresses. We focused on the case of a thin layer under compressive stress and a compliant substrate. The thin layer undergoes buckling with wavelengths and amplitudes that are proportional to its thickness, but with prefactors different from those of the theory of thin film buckling~\cite{allen,chen04}. We introduced a simplified model with a self-avoiding rod on a Winkler foundation. This model relies on linear elasticity while large strains are involved and on a thin layer approximation which is strictly valid only when its radius of curvature is larger than the thickness. Despite the shortcomings of the model, the numerical minimization of the corresponding elastic energy yielded results quantitatively closer to experiments than the classical buckling analysis, and moreover reproduced the cusp-like folds observed in experiments.

These folds are reminiscent of the convoluted shape of the brain, which might involve mechanical instabilities~\cite{brain,toro}. In fact high residual strains obtained with gels are typical of the growth of living tissues~\cite{skalak}. Therefore, the set-up that we developed might be used to mimic living tissues. Besides, much thinner layers could be prepared by spin-coating, which would yield an alternative way for micropatterning. 

\section*{Acknowledgement}

We thank Thierry Mora and Laurent Quartier for their experimental help. This study was partially supported by the Minist\`ere de la Recherche--ACI Jeunes Chercheurs and by the European Community--New and Emerging Science and Technology programs. LPS is UMR 8550 of CNRS and is associated with Universit\'e Pierre et Marie Curie (Paris VI) and Universit\'e Denis Diderot (Paris VII).

\bibliography{gels}
\bibliographystyle{unsrt}

\end{document}